\documentclass{svjour3}
\usepackage[T1]{fontenc}
\usepackage[latin9]{inputenc}
\usepackage{url}
\usepackage{amsbsy,amsmath,amssymb}
\usepackage{amstext}
\usepackage{graphicx}
\usepackage{subfig}
\usepackage{xcolor}
\newcommand{\be} {\begin{equation}}
\newcommand{\ee} {\end{equation}}
\newcommand{\bea}{\begin{eqnarray}}
\newcommand{\eea}{\end{eqnarray}}
\makeatletter
\RequirePackage{fix-cm}
\smartqed
\journalname{Celestial Mechanics and Dynamical Astronomy}
\makeatother
\begin{document}

\title{Shannon Entropy Applied to the Planar Restricted Three-Body Problem}
\author{C. Beaug\'e \and P.M. Cincotta}
\institute{C. Beaug\'e \at \emph{Instituto de Astronom\'{\i}a Te\'orica y Experimental (IATE), 
Observatorio Astron\'omico, Universidad Nacional de C\'ordoba, Argentina} \\
\email{beauge@oac.unc.edu.ar} \\
\and P.M. Cincotta \at \emph{Grupo de Caos en Sistemas Hamiltonianos, Facultad de Ciencias 
Astron\'omicas y Geof\'{\i}sicas, Universidad Nacional de La Plata and Instituto de 
Astrof\'{\i}sica de La Plata (CONICET-UNLP), Buenos Aires, Argentina}\\
\email{pmc@fcaglp.unlp.edu.ar}}
\titlerunning{Shannon Entropy Applied to the Restricted 3-Body Problem}
\authorrunning{C. Beaug\'e and P.M. Cincotta}
\date{Received: date / Accepted: date}

\maketitle

\begin{abstract}
We present a numerical study of the application of the Shannon entropy technique to the planar 
restricted three-body problem in the vicinity of first-order interior mean-motion resonances with 
the perturber. We estimate the diffusion coefficient for a series of initial conditions and 
compare the results with calculations obtained from the time evolution of the variance in the 
semimajor-axis and eccentricity plane. Adopting adequate normalization factors, both methods 
yield comparable results, although much shorter integration times are required for entropy 
calculations.

A second advantage of the use of entropy is that it is possible to obtain reliable results 
even without the use of ensembles or analysis restricted to surfaces of section or 
representative planes. This allows for a much more numerically efficient tool that may be 
incorporated into a working N-body code and applied to numerous dynamical problems in planetary 
dynamics. 

Finally, we estimate instability times for a series of initial conditions in the 2/1 and 3/2 
mean-motion resonances and compare them with times of escape obtained from directed N-body 
simulations. We find very good agreement in all cases, not only with respect to average values 
but also in their dispersion for near-by trajectories. 
\end{abstract}


\section{Introduction}

One of the most difficult questions to ask about the evolution of a planetary system is that of 
orbital stability. Simply stated, given a system of $N$ bodies of masses $m_i,\, i=1,\dots,N$
orbiting a central mass $m_0 > m_i$ under the effects of their mutual gravitational interactions,
what are the initial conditions that guarantee orbital stability for a certain time interval?
While it is possible to define Hill stability criteria in the case of the three-body problem, the
so-called Lagrange stability has proved much more difficult to tackle (see for instance H\'enon \&
Petit 1986, Gladman 1993, Sim\'o \& Stuchi 2000 and references therein). 

The KAM theory (Broer 2004 for a general description of Kolmogorov paper, Celetti \& Chierchia
2006 for the application of this theory in the context of the three body resonance in the Solar
System) and the Neckhoroshev formulation (Neckhoroshev 1977) have lead to a number of important
discoveries during the last century, including Chirkikov's resonance overlap criterion (Chirikov
1979) introduced in celestial mechanics by Wisdom (1980) and recently revisited by Ramos et
al. (2015). It is customary to difference between the strong unstable chaotic dynamics, called
Chirikov's regime, that applies when a major overlap of resonances takes place and the so-called
Neckhoroshev regime, when unstable chaotic motion is almost restricted to the very narrow chaotic
layers surronding the resonances.

Several outstanding works in planetary dynamics deal with these issues, for instance Levison et al. (1997), 
Tsiganis et al. (2005), Robutel \& Gabern (2006) for the Jupiter's Trojans asteroids;
Laskar (1989,1990, 1996,2013), Duncan \& Queen (1993), Lecar et al. (2001), Batygin \& Laughlin
(2008) concerning the stability of the Solar System; Guzzo (2005) for revealing the role of the
resonance web in the outer Solar System among hundred of relevant papers in the field.

An even more challenging question is how long an unstable planetary system will last. In other 
words, even if it is not possible to establish that a given initial condition is stable, may we
estimate how long the system will remain close to its initial configuration? Calculations of 
instability or escape times $\tau_{\rm esc}$ are possible assuming a certain chaotic diffusion 
process (normal in general) and estimating diffusion-like coefficient. Analytical methods (e.g.
Lichtemberg \& Lieberman 1983), based on a Fokker-Planck description of the dynamics in the
vicinity of separatrix of a given resonance, have been recently been employed successfully for the
GJ876 planetary systems (Batygin et al. 2015). However, they require an analytical model for the
dynamical evolution and are usually restricted to the behavior of a single resonance. However, while chaos is often associated with large instabilities, it is important to keep in mind that this is not necessarily true, the so-called ''stable chaos'' was firstly observed in the Solar System by Milani \& Nobili (1992). Thus a local exponential divergence of nearby orbits (i.e. a positive Lyapunov characteristic number) does not necessarily imply chaotic diffusion.

More general studies inevitably require a numerical approach. For instance, early works in
planetary dynamics about relations between the Lyapunov time and an ''orbital evolution time'' 
is discussed in Lecar et al. (1992). However, in more recent studies the standard procedure is 
to analyze the evolution of a fast action-type variable, say $I$, and to model the growth of its
variance as function of time. Depending on the complexity of the system, the solution $I(t)$ may
be obtained either by a discrete mapping or from the integration of the equations of motion 
in N-body simulation. Since even strongly chaotic motion is far from being ergodic in phase space,
different approaches have been considered, for instant in Cachucho et al. (2010), when studying
three body resonances the time average over a single trajectory was considered following
Chirikov's diffusion approach. On the other hand an ensemble of initial conditions were considered
in Mart\'{\i} et al. (2016), to compute space averages of action-like variables for the motion the
vicinity of the Laplace resonance in GJ876. 

Although the description of the chaotic diffusion may be obtained by any of these procedures, the
calculations become  time consuming and usually require very long-term integrations as for
instance Froeschl\'e et al. (2005), Lega et al. (2008) show in case of the slow diffusion along 
resonances in relatively simple dynamical systems or, as it was recently discussed in Cincotta et 
al. (2018) when reviewing the diffusion process in multidimensional Hamiltonian systems and 
applications to planetary dynamics.

A series of recent papers (see Giordano \& Cincotta 2018, Cincotta \& Giordano 2018, Cincotta \& 
Shevchenko 2019) have analyzed a new numerical method for studying diffusion in both, weakly and 
strong chaotic systems, based on the time evolution of the Shannon entropy. Applied to both 
discrete mappings and continuous dynamical systems it appears to constitute a valuable tool with 
which to obtain a general description of the chaotic dynamics. Moreover, it has been shown to 
yield values of the diffusion coefficients comparable to those obtained by other means, and in
shorter integration times. 

In this work we present an application of this technique to planetary systems, analyzing the 
case of the planar restricted three-body problem (R3BP) in the vicinity of the 2/1 and 3/2 
interior mean-motion resonances. We discuss how Shannon entropy calculations may be adopted to 
systems of unbounded phase space with multiple timescales and analyze both resonant and 
non-resonant trajectories. Based on a series of N-body integrations with different perturbing 
masses and initial conditions, we deduce diffusion coefficients and compare the results with 
similar 
estimations from the time evolution of the variance of the actions. We analyze how this method 
may be employed without the use of ensembles and, finally, compare the escapes times with very 
long-term numerical simulations.

\section{The 2/1 Mean-Motion Resonance}

Our dynamical system is comprised of a mass-less particle orbiting a central star of mass 
$m_0=1$ and perturbed by an exterior planet of mass $m_1 \ll m_0$. Let $a$ denote the semimajor 
axis of the mass-less body, $e$ its eccentricity, $\lambda$ the mean longitude and $\varpi$ the 
longitude of the pericenter. The same notation, with subscript one, is reserved for the 
perturber. Orbital elements are measured in a $m_0$-centric reference frame and all motion is 
restricted to the same orbital plane. 

For the current set of numerical tests we will assume the following parameters for the planet:
\be
m_1 = 2.5 \times 10^{-4} \hspace*{0.3cm} ; \hspace*{0.3cm} a_1 = 1 \hspace*{0.3cm} ; 
\hspace*{0.3cm} e_1 = 0.05
\label{eq1}
\ee
All initial angles are taken equal to zero. The adopted value for $m_1$ is slightly smaller 
than Saturn mass. 

We will consider initial conditions in the vicinity of the 2/1 mean-motion resonance (MMR), 
characterized by mean motions $n$ such that $n/n_1 \simeq (p+q)/p$ with $n_1$ the mean motion 
of the perturber and $p=q=1$. Integer $q$ is usually referred to as the order of the resonance, 
while $p$ is the degree of the commensurability. 

\begin{figure}[t]
\centering{\includegraphics*[width=1.15\columnwidth]{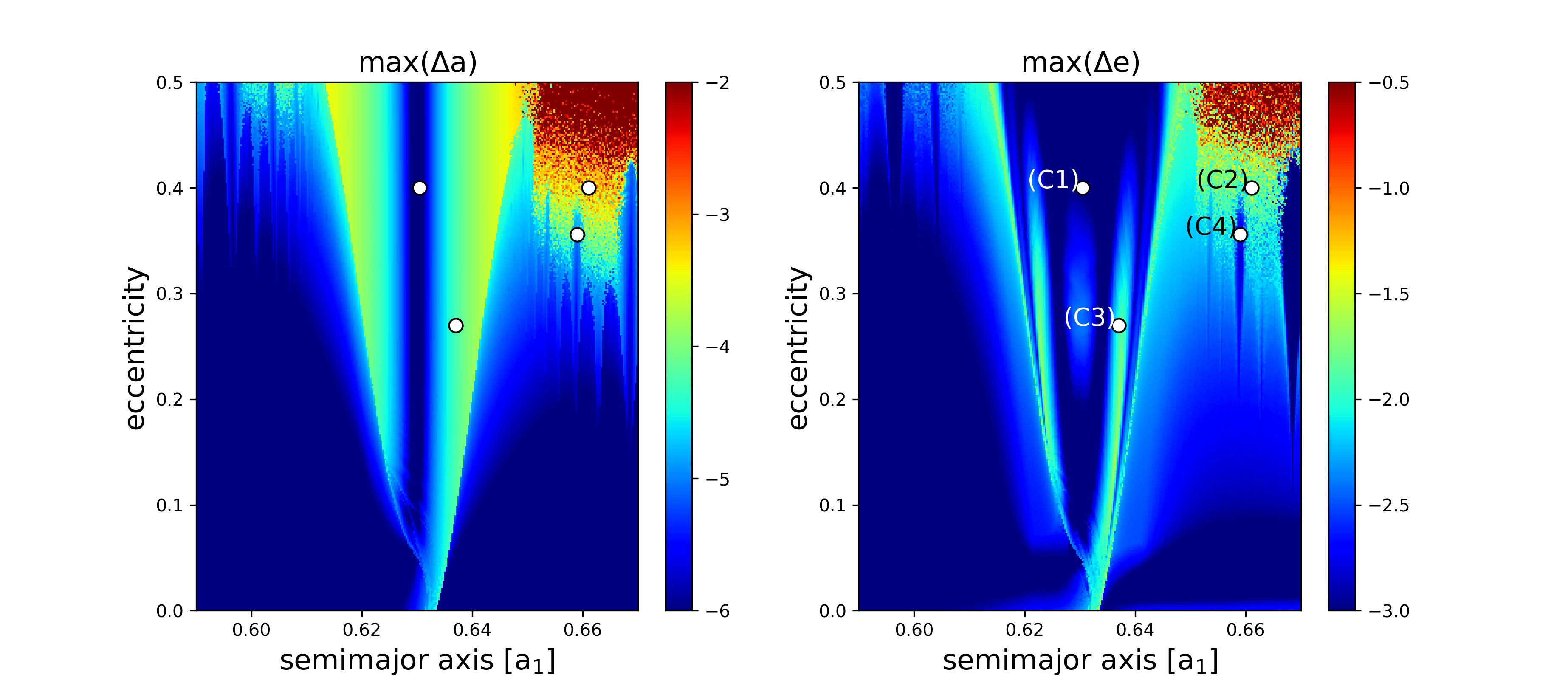}}
\caption{Dynamical maps for a grid of initial conditions in the vicinity of the interior 2/1 
MMR with a Saturn-type planet in eccentric orbit. The semimajor axis is displayed in units 
of $a_1$. The color bars accompanying each plot indicate the values of $\log{(\Delta a)}$ 
and $\log{(\Delta e)}$, respectively, associated to each color. See text for details.}
\label{fig1}
\end{figure}

Figure \ref{fig1} shows two dynamical maps resulting from the numerical integration of a set of 
mass-less bodies defining a $300 \times 300$ grid of initial conditions in the $(a,e)$ plane, 
where $a$ is taken adimensional, i.e. $a/a_1$. The equations of motion are those corresponding 
to the classical Newtonian 3-body problem, and initial values of the angles were set to zero. 
Integration time was $10^3$ years (i.e. orbits of the perturber). The color-code (form dark blue to 
orange) of the left-hand plot shows the maximum spread in semimajor axis suffered by each initial 
condition during its orbital evolution. We refer to this value as ${\rm 
max}(\Delta a)$. Similarly, the right-hand graph shows the changes experienced by the 
eccentricity: i.e. ${\rm max}(\Delta e)$.

Both plots show evidence of a forest of high-order MMR on both sides of the 2/1 commensurability
whose overlap at high eccentricities generate a broad chaotic region. Such a stochastic region 
is particularly evident for semimajor axes larger than the nominal value for the 2/1 MMR. Within 
the main resonance both maps show slightly different features. While they 
display a net of secondary resonances for $e \simeq 0.05$, the libration region in the 
left-hand plot (i.e. ${\rm max}(\Delta a)$) shows a smooth decrease in amplitude tending 
towards zero at the pericentric branch. Conversely, the plot on the right shows additional 
structures related to secular resonances inside the libration domain. An additional moderate
${\rm max}(\Delta e)$ central region is also noted for $e\ \simeq 0.3$. 

\begin{figure}[t]
\centering{\includegraphics*[width=0.9\columnwidth]{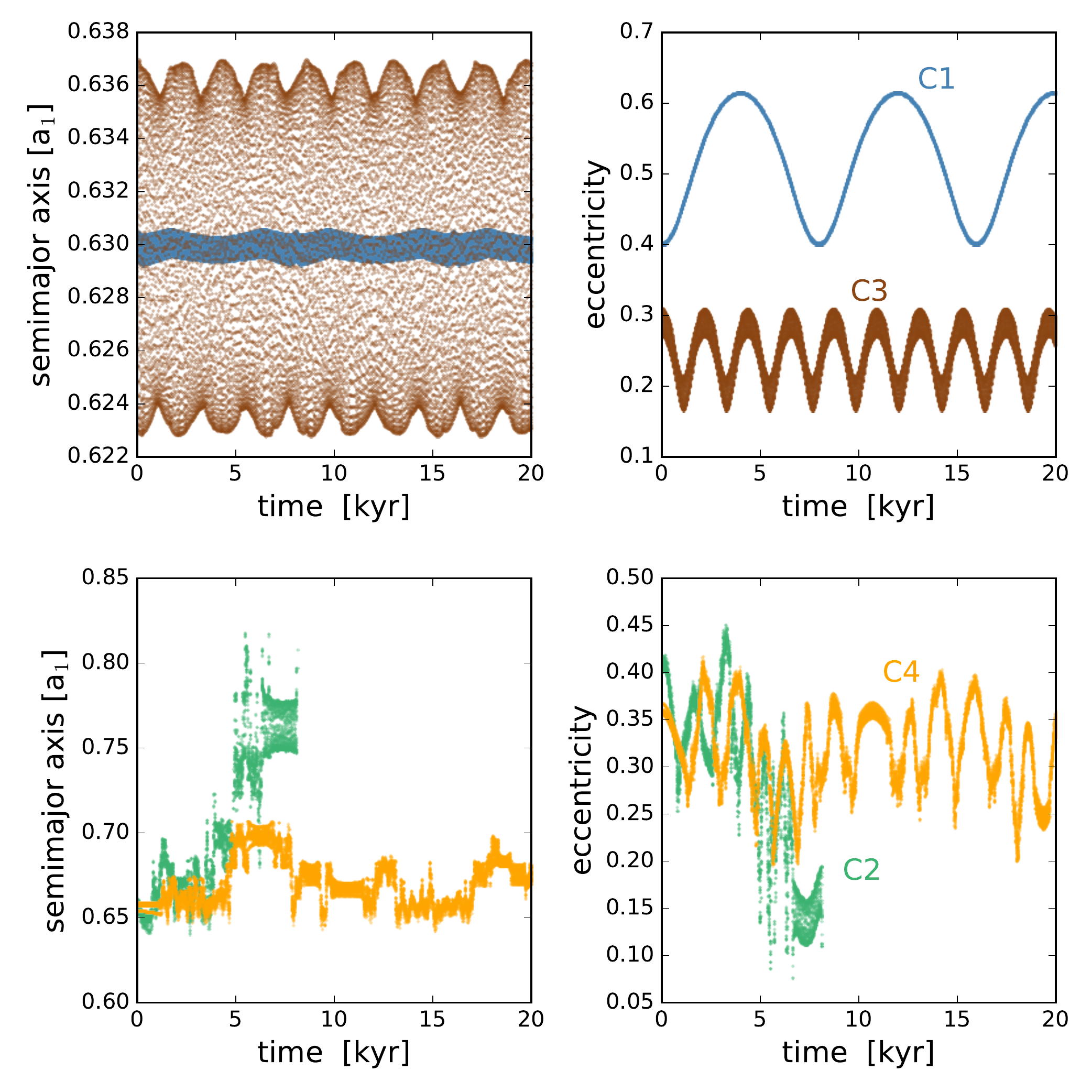}}
\caption{Numerical integration of four mass-less particles with initial conditions given by 
(\ref{eq2}). Resonant orbits are shown on the top frames, with (C1) in blue and (C3) in brown.
Trajectories initially outside the 2/1 MMR are presented in the lower plots, with (C2) in 
green and (C4) in light orange.}
\label{fig2}
\end{figure}

\section{Individual Runs}

From these maps we chose four initial conditions, indicated in Figure \ref{fig1} by white 
circles and numbered in the right-hand plot. The semimajor axes and eccentricities are:
\be
\begin{split}
{\rm (C1)}: \hspace*{0.3cm} & a = 0.6305 \hspace*{0.3cm} ; \hspace*{0.3cm} e = 0.4 \\
{\rm (C2)}: \hspace*{0.3cm} & a = 0.6611 \hspace*{0.3cm} ; \hspace*{0.3cm} e = 0.4 \\
{\rm (C3)}: \hspace*{0.3cm} & a = 0.6370 \hspace*{0.3cm} ; \hspace*{0.3cm} e = 0.27 \\
{\rm (C4)}: \hspace*{0.3cm} & a = 0.6590 \hspace*{0.3cm} ; \hspace*{0.3cm} e = 0.356
\label{eq2}
\end{split}
\ee
Initial conditions (C1) and (C3) lie within the libration region of the 2/1 MMR, the first  
close to the pericentric branch and the second in the vicinity of the secular resonance 
generated inside the resonant domain. The other two initial conditions were chosen outside the 
commensurability in a region dominated by long-term chaotic motion. While (C2) exhibits 
short-term chaoticity and is eventually ejected from the system, (C4) initially lies in a 
high-order MMR and displays a seemingly regular orbit during a few $10^3$ yrs, and the chaotic 
nature of its motion is only noticeable for longer integrations (see below). 

\begin{figure}[t]
\sidecaption
\includegraphics*[width=0.70\columnwidth]{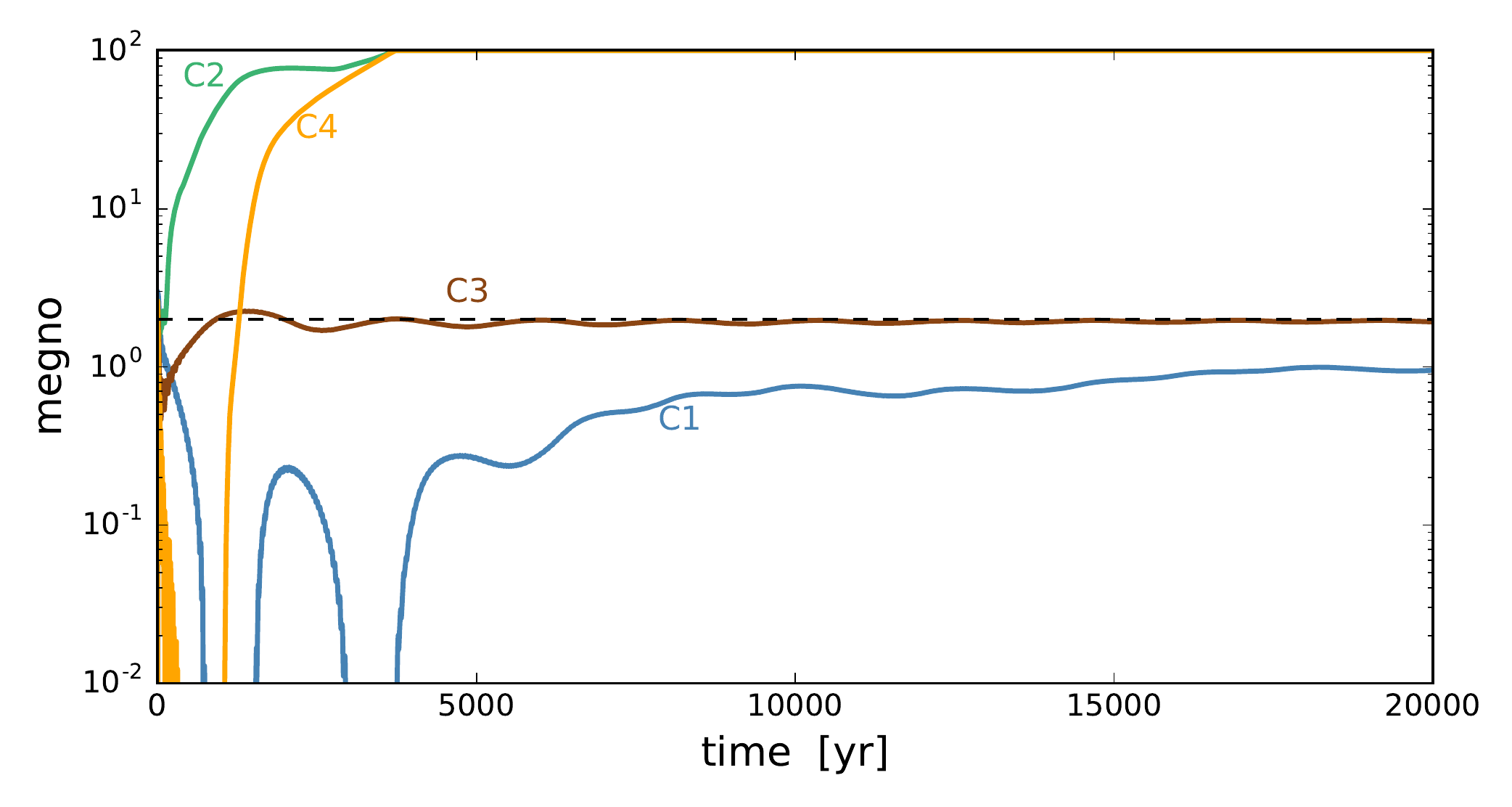}
\caption{Time evolution of the MEGNO chaos indicator for the same four initial conditions 
presented in Figure \ref{fig2}. The dashed horizontal black line corresponds to the threshold
MEGNO value $2$.}
\label{fig3}
\end{figure}

\subsection{Orbital Evolution}

All four initial conditions were numerically integrated with a Bulrisch-Stoer code for $2 
\times 10^4$ orbits of the perturber. Figure \ref{fig2} shows the evolution of the semimajor 
axis (left-hand frames) and the eccentricity (right-hand plots). The two upper graphs show 
results for (C1) in blue and for (C3) in brown, both initial conditions within the 2/1 
resonance domain. In any case, the orbital elements display seemingly quasi-periodic motion, 
which is confirmed by the calculation of MEGNO (Cincotta \& Sim\'o 2000, Cincotta et. al 2003, 
Cincotta \& Giordano 2016), as presented in Figure \ref{fig3}. Notice that the value of MEGNO 
for (C3) appears to converge to a value close to 2, indicative of quasi-periodic regular 
motion, while that of (C1) remains below this limit throughout the integration time-span, 
indicative of a motion in a lower-dimensional torus. 

Returning to Figure {\ref{fig1}, the evolution of the semimajor axis is dominated by 
short-period terms with very low amplitude in the case of (C1) and a much larger periodic 
variation in the case of (C3). The changes in the eccentricities, however, are primarily 
defined by the resonant and secular perturbing terms and have much larger periods, of the order 
of $10^3$ yrs. The large-amplitude oscillation perceived for (C1) is mainly driven by the 
forced eccentricity term and should decrease to zero as $e_1 \rightarrow 0$. 

The two lower plots in Figure {\ref{fig1} correspond to those initial conditions outside the 
2/1 MMR domain; (C2) is drawn in green while (C4) in light orange. Both are extremely chaotic 
(indicated in Figure \ref{fig3} by large MEGNO values), and the chaotic diffusion is exhibited 
primarily in the semimajor axis and not in the eccentricity. While the eccentricities show 
irregular variations, these appear bounded and with a limited long-term trend. The semimajor axes, 
on the other hand, experience erratic alternations between different high-order resonances that 
lead to a chaotic diffusion in the action space. 

Although (C2) and (C4) are both highly irregular, their orbital evolution show significant 
differences, especially for short time spans. Orbit (C2) lies in a global chaotic sea and is 
rapidly ejected from the system; its MEGNO indicator grows monotonically with time almost from 
the beginning of the simulation. Initial condition (C4), on the other hand, was placed in what 
appeared to be a regular island inside a high-order MMR. Its short-term evolution is thus 
regular and with a very low value of MEGNO. However, after a few $10^3$ years it is removed 
from the commensurability, enters in the chaotic domain and begins to diffuse in the semimajor 
axis space. The value of MEGNO grows quickly, although the particle is never ejected from the 
system, at least during the integrated time-span.

\subsection{Shannon Entropy}

The application of Shannon entropy as an indicator of chaos and diffusion in phase space of 
dynamical systems stems from several years back (see Giordano \& Cincotta 2018, Cincotta \& 
Giordano 2018 and references therein). This tool was proposed to provide a measure of the extent 
of the instability region in action space as well as a good estimate of the diffusion rate.
Moreover, in Cincotta \& Giordano (2018) and Cincotta \& Shevchenko (2019) it was shown the
efficiency of the Shannon entropy to detect correlations among the state variables, even when
they are extremely weak. However, all these applications are restricted to relatively simple 
symplectic maps or to the Arnold Hamiltonian (Arnold 1964, Chirikov 1979).   

The numerical calculation of Shannon entropy, action variances and diffusion rates (e.g. 
Mart\'{\i} et al. 2016, Cincotta et al. 2018, Giordano \& Cincotta 2018) usually employ two 
different techniques to reduce the numerical noise and increase the precision. In this direction
each initial condition is represented by an ensemble of particles with a very small initial 
dispersion in orbital elements. Additionally, the relevant quantities are evaluated in a given
representative plane, i.e. when the angles acquire values very close to those corresponding to 
$t=0$. In particular, the computation of the entropy requires a partition the action space,
in the present application, it implies a grid of bi-dimensional cells defined on a given
region of the $(a,e)$ plane that cover this domain of the action space (see below).

Although these practices are relatively simple to adopt in the case of mappings or simple 
continuous dynamical systems, they prove much more difficult in more complex systems 
such as the one discussed here. A secular timescale of the order of $2\pi/g \simeq 10^3$ years 
implies that the integration time of each initial condition must extend to values close to the 
age of the Solar System if we desire a total number of orbital points significantly larger than 
the number of cells in each partition. This makes any large-scale analysis of the system 
prohibitively expensive in terms of CPU usage. Moreover, if the orbit is extremely chaotic, 
intersections with the representative plane may be very difficult to obtain, as was shown by 
Mart\'{\i} et al. (2016) in the case of Gliese-876 or by Maffione et al. (2016) for halo stars 
in a neighborhood of the Sun.

In principle, however, it is not necessary to restrict the points of the trajectory to
any reference plane as for instance the one defined by Fig. \ref{fig1}, where all the angles 
take the very same value. For example, let us assume a generic 2 dof near-integrable dynamical 
system, written in terms of action-angles variables $(J_1,J_2,\theta_1,\theta_2)$ of the 
unperturbed Hamiltonian. A regular trajectory will be characterized by a single point or a 
small curve in $(J_1,J_2)$ when restricting the motion to the representative plane defined by 
$(\theta_1,\theta_2) = (\theta_{1_0},\theta_{2_0})$, and thus easily identifiable as a strongly
localized distribution in any partition of the action plane that contain at least more than 
one element. Chaotic motion, on the other hand,  will appear in general as a two-dimensional 
region of the action space. If the chaos is not local (i.e. bounded) but global, then the 
area of this region of the action space may increase with time according to the associated 
diffusion rate. 

If we now analyze the projection of the full phase space trajectory onto the action space, not 
restricted to any representative plane, all quasiperiodic orbits will define a bounded 
two-dimensional surface and thus the area of the action space covered by regular motion would be 
always confined to a small subspace. Chaotic orbits would also cover some surface of phase space 
but of increasing area. That is chaotic diffusion will induce a change in the area covered by 
the orbit and for timescales larger than that corresponding to the lowest frequency of the 
system, the diffusion could be detected.  

Therefore in the following discussion we will generalize the formulation given in Cincotta \& 
Shevchenko (2019), Giordano \& Cincotta (2018) for the  Shannon entropy as a tool to 
characterize the nature of the motion without restricting the motion to any reference plane, 
however in Subsection \ref{planes-ensembles} a more detailed analysis is provided.

The restricted 3-body problem has 3 distinct degrees of freedom, each with its own particular 
timescale. The highest frequency is associated to the synodic angle $Q$ and has a period of the 
order of the orbital periods ($\simeq 1$ year in our simulations). On the opposite end, the 
lowest frequency is defined by the secular evolution of the longitude of pericenter $\varpi$ 
and, in our studies, is of the order of several $10^3$ years. The remaining degree of freedom is 
dictated by the critical angle of whichever MMR dominates the resonant dynamics, if any. Its 
frequency depends on the commensurability, mean eccentricity and the proximity of the initial 
condition to the center of the libration domain. 

Since we expect chaotic diffusion to be particularly noticeable in the actions, we determined 
the entropy in the semimajor axis and eccentricity $(a/a_1,e)$ plane. This is not the only 
option; we performed additional tests using canonical variables (e.g. modified Delaunay 
actions). No significant differences were found, indicating that the particular choice of 
action-like variables should be fairly robust. For simplicity, hereafter we will refer to 
$(a/a_1,e)$ as the {\it action plane} even though both orbital elements are not truly action 
variables of the system.

Our first step is to define boundaries for the trajectories in the $(a/a_1,e)$ plane. Although 
angles are naturally bounded in the range $[0,2\pi)$, the actions prove a greater challenge. We 
finally opted, after different trials,  for defining the limits as $[a_{\rm min},a_{\rm max}] 
\times [e_{\rm min},e_{\rm max}]$ with
\be
\begin{split}
a_{\rm min} = a_0 - \delta_a  \hspace*{0.5cm} ;& \hspace*{0.5cm} a_{\rm max} = a_0 + \delta_a \\
e_{\rm min} = e_0 - \delta_e  \hspace*{0.5cm} ;& \hspace*{0.5cm} e_{\rm max} = e_0 + \delta_e,
\end{split}
\label{eq3}
\ee
where $a_0$ and $e_0$ are the initial values of the semimajor axis and eccentricity, while 
$\delta_a$ and $\delta_e$ are the sizes of the box in orbital elements where the partition will 
be introduced. We chose $\delta_a = 0.1 a_1$ and $\delta_e = 0.3$ since these were typical 
amplitudes found in initial runs for stable and bounded chaotic trajectories; however the 
results were not very sensitive to changes in the limits. We used periodic boundary conditions 
when one or both of the actions exceeded the limits of the box. This allowed us to follow the 
evolution of the initial condition regardless the size of its excursion in semimajor axis and/or 
eccentricity, while at the same time preserving an adequate resolution in the case of very 
regular small-amplitude oscillations.

The plane of the actions was divided into a rectangular grid of $r = r_a \times r_e$ cells. In 
most of our runs we adopted $r_a = r_e = 400$, leading to a total of $1.6 \times 10^5$ cells 
per plane. Although we also did several tests with larger and smaller values, no large-scale 
differences were found in the results. In the next section we will discuss the sensitivity of 
the entropy and diffusion coefficient with $r$.

At any given time $t$ of the integration, let us denote by $N = t/h$ the number of orbital 
points of a given trajectory $\gamma$, with $h$ the output step. In our case, we chose $h = 
10^{-2}$ years, corresponding to one-hundredth of the orbital period of the perturber. We denote 
with  $n_k$ the number of times $\gamma$ fell in the $k$-th cell of the partition. We can then 
calculate the Shannon entropy as (see for example Cincotta \& Shevchenko 2019 for details):
\be
S(\gamma,N) = \ln{N} - \frac{1}{N} \sum_{k=1}^{r} n_k \ln{(n_k)},
\label{eq4}
\ee
which in fact depends on the partition. It is simple to show that $S$ presents to extreme 
values; its minimum, $S=0$, when $n_k=\delta_{ik}$, that is all the action values lie in a 
single cell, the $i$ cell. On the other hand, when $n_k=N/r$ (ergodic motion) the entropy takes 
its maximum, $S = \ln r$. 

In the case of a nearly ergodic orbit $\gamma_e$ that cover  $r_0\le r$ cells, the distribution 
of orbital points in the grid would present small deviations from the mean value $N/r_0$ and it 
is possible to relate the entropy with occupied cells $r_0$, by
\be
S(\gamma_e,N) \simeq \ln{(r_0)} - \frac{r_0}{2N} R
\label{eq5}
\ee
where $R$ is a constant parameter that depends on the dynamics that defines the distribution of 
points. If the latter is Poissonian (completely random motion), $R=1$. As the number of points 
becomes much larger than the number of occupied cells, for any chaotic trajectory $\gamma$, we 
may take $S(\gamma,N)\approx \ln{(r_0)}$ as an approximate value for the entropy.

Figure \ref{fig4} shows the normalized entropy, $S/\ln r$ (left-hand plot) and the natural 
logarithm of the number of occupied cells (right-hand frame). The color code employed for each 
initial condition is the same defined in previous figure, where now the total integration time 
was extended to $T=10^5$ years. Each initial condition was represented by an ensemble of 
$N_{\rm part}=100$ ghost particles distributed randomly in semimajor axis and eccentricity 
around the nominal values $a_0$ and $e_0$ with a maximum range $\Delta a/a_0 = \Delta e = 
10^{-7}$. The use of ensembles, as mentioned, smooths oscillations in the orbital evolution due 
to microscopic changes in the starting values, as well as increasing the number of orbital 
points per unit time. In fact, when working with ensembles the value of $N$ in expressions 
(\ref{eq4}) and (\ref{eq5}) should be replaced by $\bar{N}=N_{\rm part} \times N$. 

Since part of the original ensembles corresponding to chaotic orbits were ejected from the 
system during the run, we re-introduced each eliminated body back into the simulation with a new 
initial condition chosen within the spatial limits of the ensemble as a reflection. Although 
this technique eliminates the possibility of following the orbital evolution of individual 
particles, we found that it does not introduce any significant noise in the estimations of the 
entropy while allowing for integrations well beyond the typical escape time of any region of the 
phase space.

\begin{figure}[t]
\includegraphics*[width=0.99\columnwidth]{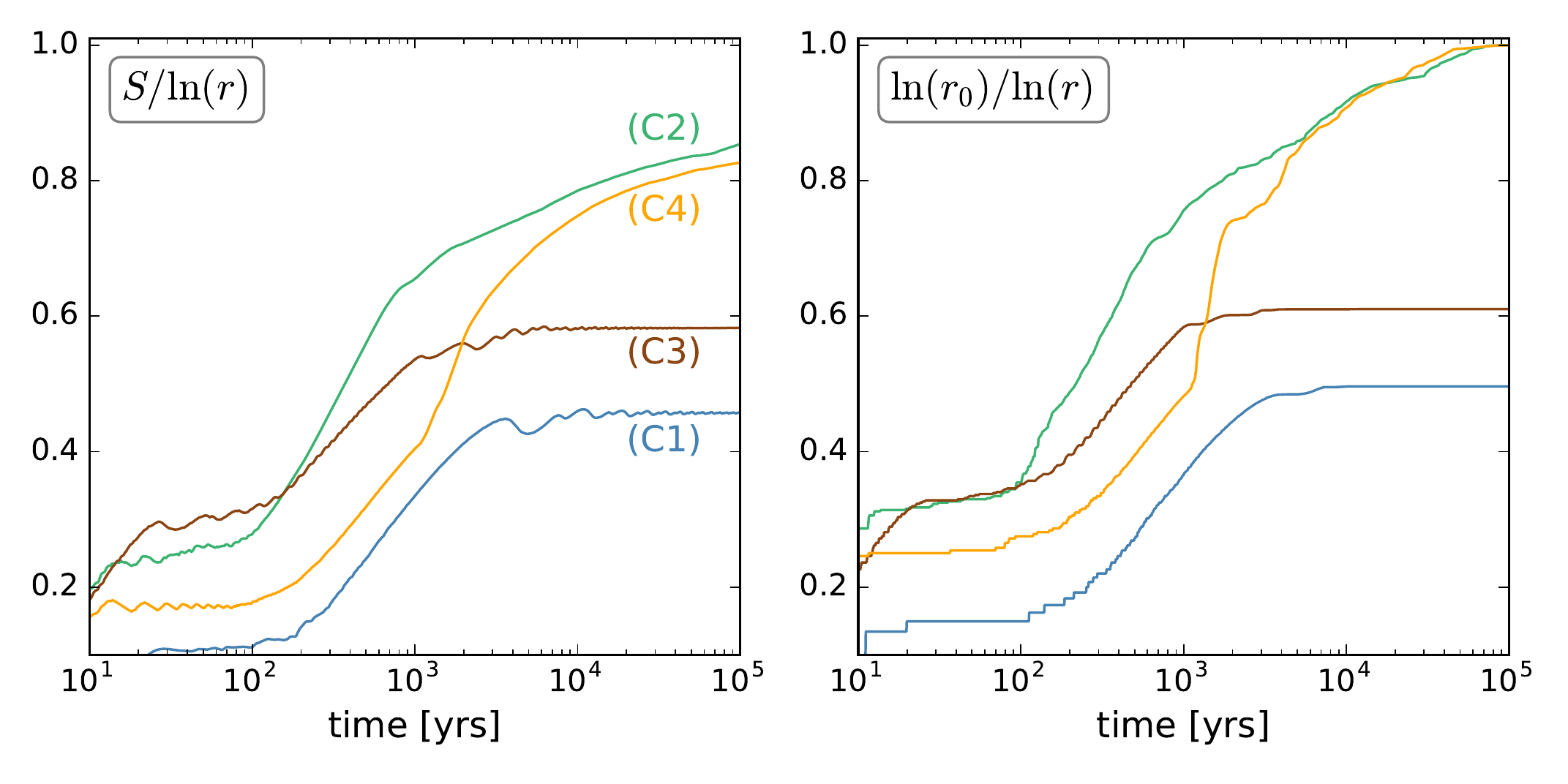}
\caption{{\bf Left:} Time evolution of the Shannon entropy calculated from ensembles of 
$N_{\rm part}= 100$ ghost particles around each initial condition (C1)-(C4). {\bf Right:} 
Natural logarithm of the number of occupied cells. Total integration time was extended to $10^5$ 
years. All values are given in units of $\ln{(r)}$.}
\label{fig4}
\end{figure}

The behavior of both $S$ and $\ln{(r_0)}$ show similar general trends. After $t \sim 10^4$ yrs, 
regular orbits associated to initial conditions (C1) and (C3) level off at maximum values much 
smaller than $\ln{(r)}$ indicating that both the number of occupied cells and the distribution 
of $n_k$ within them stabilized, the orbital points are confined to a small region of the action 
space. The final values of $\ln{(r_0)}$ is proportional to the libration amplitude of each 
trajectory within the 2/1 MMR; a lower number corresponding to (C1) initially near the 
pericentric branch, and a higher value in the case of the initial condition (C3) close to the 
separatrix. In other words, the plateau observed for both (C1) and (C3) are due to a motion in a 
distorted torus and indicative of regular motion. In contrast, $r_0$ for both chaotic orbits 
(C2) and (C4) continues growing until saturating the partition near the end of the run. Since 
the distribution of the orbital points is not uniform, the entropy $S$ continues to increase 
even after the available cells are completed.

These plots show three distinct timescales, one related to the synodic period {\bf ($\simeq 10^0-10^1$ years)}, a second defined by the secular dynamics and with period of the order of $\tau_g \sim 10^3$ years, plus a third timescale associated to chaotic diffusion and unrelated to periodic motion. Regardless of the initial condition, for $t \ll \tau_g$ the amplitude of both $a$ 
and $e$ are dictated by short-period perturbations and the trajectory remains relatively 
confined to a small number of cells. However, as $t \rightarrow \tau_g$, the large amplitude 
increase in the eccentricity from secular perturbations becomes significant, and the entropy 
grows accordingly reaching more or less constant slopes for $ t > \tau_g$. This behavior 
indicates that the minimum time-span of the numerical integration must cover at least a few 
secular periods of the system.

\subsection{Estimation of the Diffusion Rates from the Variance}

Numerical estimations of the rates of chaotic diffusion in the action plane may be obtained 
studying the time evolution of the variance
\be
{\rm Var} (I) = \frac{1}{\bar{N}} \sum_{k=1}^{\bar{N}} I_k^2 - \mu^2,
\label{eq6}
\ee
where $I_k \equiv I(t_k) = [(a(t_k)/a_1)^2 + e(t_k)^2]^{1/2}$ is the Euclidean metric covered 
by the trajectory in the action plane, $t_k$ is its value at time $t=t_k$, 
$\bar{N}=N_{\mathrm{part}}\times N$ and $\mu$ is the average value at the same instant and, i.e.
\be
\mu = \frac{1}{\bar{N}} \sum_{k=1}^{\bar{N}} I_k .
\label{eq7}
\ee
Thus defined the variance includes both time and space average. Alternatively, it is customary 
to compute the ensemble variance averaging over $N_{\rm part}$ instead of $\bar{N}$. Assuming 
nearly normal diffusion, we can then estimate the diffusion coefficient $D_{\rm var}$ as 
the mean time derivative of the variance. Indeed, as it was discussed for instance in 
Cincotta et al. (2018), in near-integrable Hamiltonians the variance of any fast action-like 
variable for comparatively short motion times does not scale linearly with time and thus, the 
derivation of a diffusion coefficient in the standard way by means of a linear fit in general 
does not work.

\begin{figure}[t]
\includegraphics*[width=0.99\columnwidth]{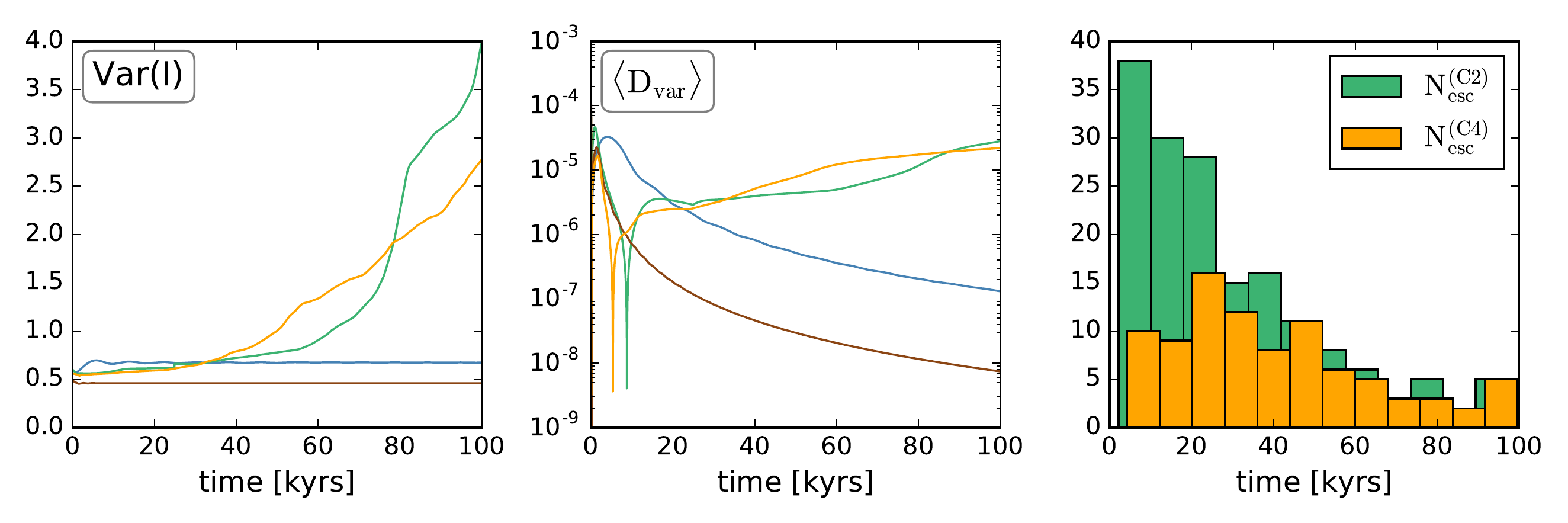}
\caption{{\bf Left:} Variance of $I = [(a/a_1)^2 + e^2]^{1/2}$, calculated for ensembles 
around all four nominal initial conditions, as function of time up to $T = 10^5$ years. 
The same color code is used as in the previous figure. {\bf Center:} Normal diffusion 
coefficient $D_{\rm var}$, estimated from the mean time derivative of the variance. {\bf 
Right:} Histograms representing the escape times of members of the ensembles associated to (C2) 
and (C4).}
\label{fig5}
\end{figure}

Figure \ref{fig5} presents the results for the same numerical simulations described before. The 
color code follows the same pattern as in previous figures. The left-hand frame shows the time 
evolution of ${\rm Var} (I)$ for each initial condition. While the ensembles associated to 
regular orbits yield variances that rapidly stabilize around values close to zero, those 
corresponding to the highly chaotic ensembles exhibit monotonic growths. The slopes, however, do 
not approach constant values for large integration times, but seem to exhibit different speeds 
of diffusion at different time intervals. This behavior could be related to temporary capture 
(stickiness) in high-order commensurabilities, and thus representative of a diffusion that 
occurs in a phase space that is not free of remnant structures. Even so, during the last few 
$10^4$ years the curves of both (C2) and (C4) show similar trends.

The middle plot shows the estimation of the diffusion coefficient $\langle D_{\rm var} \rangle$ 
where the derivative of the variance was averaged over time. While the values for the regular 
orbits tend to zero, those calculated for (C2) and (C4) appear to approach asymptotic values of 
the order of $10^{-5}-10^{-4}$, in units of 1/yr. In theory, the inverse of the diffusion 
coefficient should be indicative of the time required by a certain initial condition to change 
its actions by order of unity; in other words, the typical escape time from the system. The 
histograms displayed in the right-hand graph shows the distribution of the escape times of 
members of the chaotic ensembles obtained by a direct N-body integration. Although both 
distributions are not equivalent, most of the ejection times of particles from the system 
appears to take place between $10^{4}-10^{5}$ years, thus confirming, at least in the 
order-of-magnitude range, the values deduced from the estimation of the diffusion coefficient.

\subsection{Diffusion Rates from the Entropy}

As discussed by Giordano \& Cincotta (2018) and Cincotta \& Shevchenko (2019), the time 
evolution of the Shannon entropy $S$ may be used to estimate the rate of chaotic diffusion along 
a given plane, for example the one defined by the orbital elements $(a,e)$. Given a number of 
orbital points $\bar{N}$ such that $r_0/\bar{N}\ll 1$, from (\ref{eq5}) we can then 
approximately relate the rate of change in $S$ with $r_0$ through
\be
\frac{dS}{dt} \simeq \frac{1}{r_0} \frac{d r_0}{dt} .
\label{eq8}
\ee
This expression was derived in the above mentioned works assuming a nearly uniform distribution
of orbital points within the occupied cells of the partition. This condition is expected to be 
a good approximation of the dynamics in the case of highly chaotic trajectories, and provided 
the integration time is sufficiently long compared with the timescales of the perturbations.

We next assume a direct relation between the variance ${\rm Var} (I)$ and the number of 
occupied cells, such that 
\be
\frac{r_0}{r} \simeq \frac{{\rm Var} (I)}{({I}_{\rm max}-{I}_{\rm min})^2} ,
\label{eq9}
\ee
where $(I_{\rm max}-I_{\rm min})$ is the total interval of the action $I$ observed for 
a given ensemble during its dynamical evolution. Equation (\ref{eq9}) simply states that the 
changes in the orbital elements are proportional to the variation in the number of occupied 
cells, where both quantities are proper normalized. The reader is referred to Giordano \& 
Cincotta (2018) and Cincotta \& Shevchenko (2019) for a more detailed discussion.

Differentiating (\ref{eq9}) with respect to time, we may write
\be
\frac{d}{dt} {\rm Var} (I) = \frac{({I}_{\rm max}-{I}_{\rm min})^2}{r} \frac{d r_0}{dt} 
\simeq \frac{({I}_{\rm max}-{I}_{\rm min})^2}{r} r_0 \frac{d S}{dt} ,
\label{eq10}
\ee
where the last equality is obtained through approximation (\ref{eq8}). In fact, after a series 
of test runs, we have found that the time variation of the entropy is actually a better 
indicator of diffusion than the derivative of $r_0$. A probable explanation may lie in the 
fact that an increase in the variance ${\rm Var} (I)$ is not only tied to the number of 
occupied cells but also to the change in their population. The entropy $S$ keeps track of both 
these dynamical signatures, while $r_0$ only detects the former.

Assuming nearly normal diffusion in action space we can relate the change of the variance with 
time as $\delta {\rm Var} (I) \simeq D \delta t$, from which we can estimate the diffusion 
coefficient as:
\be
D \simeq \Lambda \, r_0 \frac{d S}{dt}  ,
\label{eq11}
\ee
where the scale-ratio $\Lambda$ between action plane and partition is given by
\be
\Lambda = \frac{({I}_{\rm max}-{I}_{\rm min})^2}{r} .
\label{eq12}
\ee
Note that $\Lambda$ is not a constant parameter but a function of the dynamical evolution of the 
system. Its magnitude is defined by the excursions of the given trajectory in the action space 
during the numerical simulation as well as by the size of the partition. 

\begin{figure}[t]
\sidecaption
\includegraphics*[width=0.7\columnwidth]{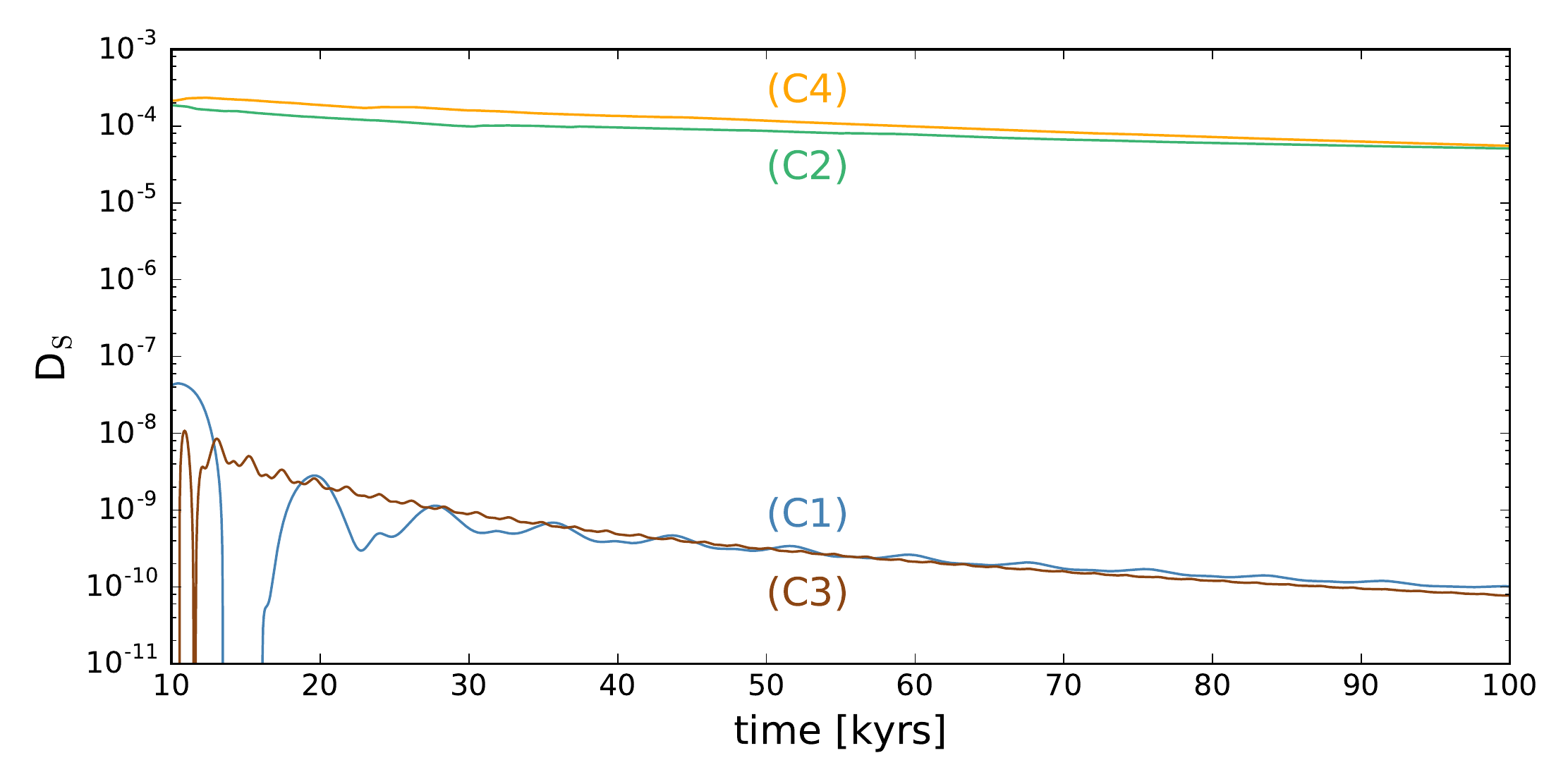}
\caption{Estimation of the diffusion coefficient (dubbed $D_S$) from the averaged time 
derivative of the Shannon entropy as given by equation (\ref{eq11}).}
\label{fig6}
\end{figure}

Figure \ref{fig6} shows the estimations of $D_S$ as a function of time for all four initial 
conditions (C1) through (C4). The overall behavior displays a very good agreement with $D_{\rm 
var}$ presented in Figure \ref{fig5}, both for regular and for chaotic orbits. While the values 
of $D_S$ for the chaotic initial conditions (C2) and (C4) remain more or less constant, those 
corresponding to $D_S$ for (C1) and (C3) rapidly fall to values close to zero, even for short 
integration times. Comparing these results with those of $D_{\rm var}$, they seem to indicate 
that diffusion estimates from Shannon entropy are possible even from short-term integrations, a 
result that was already found in Giordano \& Cincotta (2018) for toy dynamical models.

The most noticeable difference between $D_S$ and $D_{\rm var}$ may be the long-term trend 
observed for the chaotic trajectories. While the diffusion coefficient estimated from the 
variance seems to exhibit a secular growth in magnitude, the opposite occurs for $D_S$. 
Follow-up integrations extended to $10^6$ years indicate that none of these indicators reach 
constant or asymptotic values, but always display long-term fluctuations as function of time. 
The amplitude of these fluctuations is more or less the same for both indicators, limited 
between values of $10^{-5}$ and $10^{-4}$.

\subsection{On the representative planes and ensembles}
\label{planes-ensembles}

As we have already discussed, it seems not necessary to restrict the motion to any 
representative plane to compute the entropy as it was done in Giordano \& Cincotta (2018). 
Indeed, the time-evolution of $S$ given in Figure \ref{fig4} allow us to understand the global 
behavior of the entropy in this case. The growth of $S$ up to the secular timescale is primarily 
fueled by the long-period orbital variations, while the systematic increase in the entropy 
observed for longer time-spans is caused by chaotic diffusion. In other words, as long as the 
total integration time-spans least a few secular timescales, chaos and diffusion should still 
be observable and possible to estimate in the plane of the actions even without a reduction to 
any representative plane. 

The other technical aspect we will analyze is the use of ensembles, where the dynamical 
evolution of a given initial condition is studied following the trajectory of $N_{\rm part}$ 
ghost particles originally distributed in a very small region around the nominal values. The 
adopted $N_{\rm part}$ depends on the system, but usually takes values between $10^2-10^3$ (e.g. 
Cincotta et al. 2014, Mart\'{\i} et al. 2016, Cincotta et al, 2018). The main problem with 
ensembles is, once again, CPU time. The same computer time required to integrate 100 ghost 
particles could be employed to extend the simulation of a single initial condition by two orders 
of magnitude or, similarly, map a significant region of the action space. We therefore wish to 
study how the calculation of the entropy depends on $N_{\rm part}$ and whether working without 
ensembles leads to similar results. 

\begin{figure}[t]
\includegraphics*[width=0.99\columnwidth]{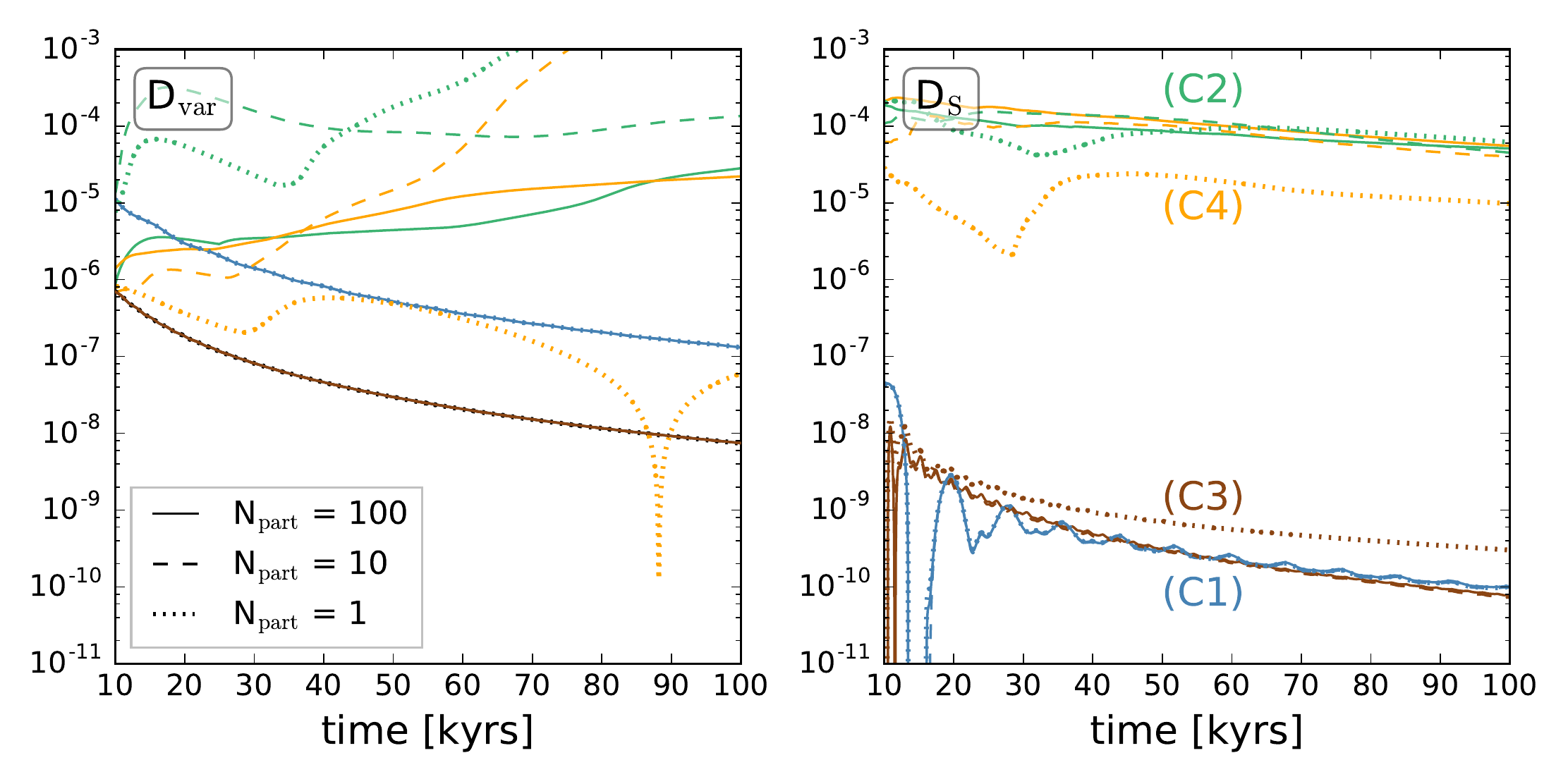}
\caption{Time evolution of the diffusion coefficient $D_{\rm var}$ (left) and $D_S$ (right) for 
all four initial conditions (C1)-(C4) and employing three different number of particles for the 
ensembles. The case $N_{\rm part} = 1$ corresponds to a single particle per initial condition.}
\label{fig7}
\end{figure}

Figure \ref{fig7} compares estimations of the diffusion coefficient for (C1)-(C4) adopting 
different number of particles $N_{\rm part}$ around each initial condition. Continuous curves 
reproduce our previous results with 100 ghost particles while dashed lines present results for 
$N_{\rm part} = 10$. Finally, dotted lines show values obtained without the use of ensembles. 

The left-hand frame shows that the variance ${\rm Var}(I)$ appears to be extremely sensitive to 
the number of ghost particles and no credible estimations are possible without ensembles. This 
behavior is probably an indication that the phase space does not satisfy the ergodicity 
condition: time averages are not equal to ensemble averages. In particular for $N_{\rm part} = 
1$, only the time average is involved in the variance. Consequently, estimations of the 
diffusion rate from the variance always seem to require large ensembles. 

A different story is observed in the right-hand plot, where the diffusion rate was estimated 
from the Shannon entropy. While some dispersion is noted, particularly for single particles 
in (C3) and (C4), the overall trends are preserved and the qualitative nature of the dynamics 
appears to be correctly for all values of $N_{\rm part}$. The case of (C4) is particularly 
interesting. The rapid decrease in $D_S$ observed in the early stages of evolution for $N_{\rm 
part}=1$ is related to the time spent by the particle inside the original high-order MMR. The 
dynamics therefore appears regular and no significant diffusion seems to take place. As the 
secular evolution drives the body beyond the limits of the commensurability into the chaotic 
domain, the diffusion increases. The time-evolution of $D_S$ after this point follows closely 
the behavior found using ensembles. 

The results for (C4) raise the question whether the sensitivity observed in $D_S$ with respect 
to $N_{\rm part}$ is due to an increasing inaccuracy of the diffusion coefficient with smaller 
ensembles, or a reflection of local dynamical behaviors that are blurred when following the 
evolution of finite (even very small) regions of the phase space. We will address this issue in 
forthcoming sections by comparing the escape times predicted by $D_S$ with long-term numerical 
simulations.

In any case, since the computed values of the entropy for a given trajectory only depend on the 
calculation of the numerical measure of the elements of the partition, $S$ as well as its time 
derivative should be almost insensitive to the use of ensembles provided that $N=t/h \gg r$.

\section{Initial Conditions in a Line Segment}

Let us now consider a set of 400 initial conditions with semimajor axes $a/a_1 \in [0.59,0.67]$ 
and eccentricity $e = 0.4$. All other orbital elements are the same as discussed previously. 
We numerically integrated this set for a total time-span of $T=10^5$ years and calculated both 
the diffusion coefficient $D_S$ as determined from the entropy, as well as the value of the 
MEGNO chaos indicator. Results are shown in Figure \ref{fig8}. 

\begin{figure}[t]
\includegraphics*[width=0.99\columnwidth]{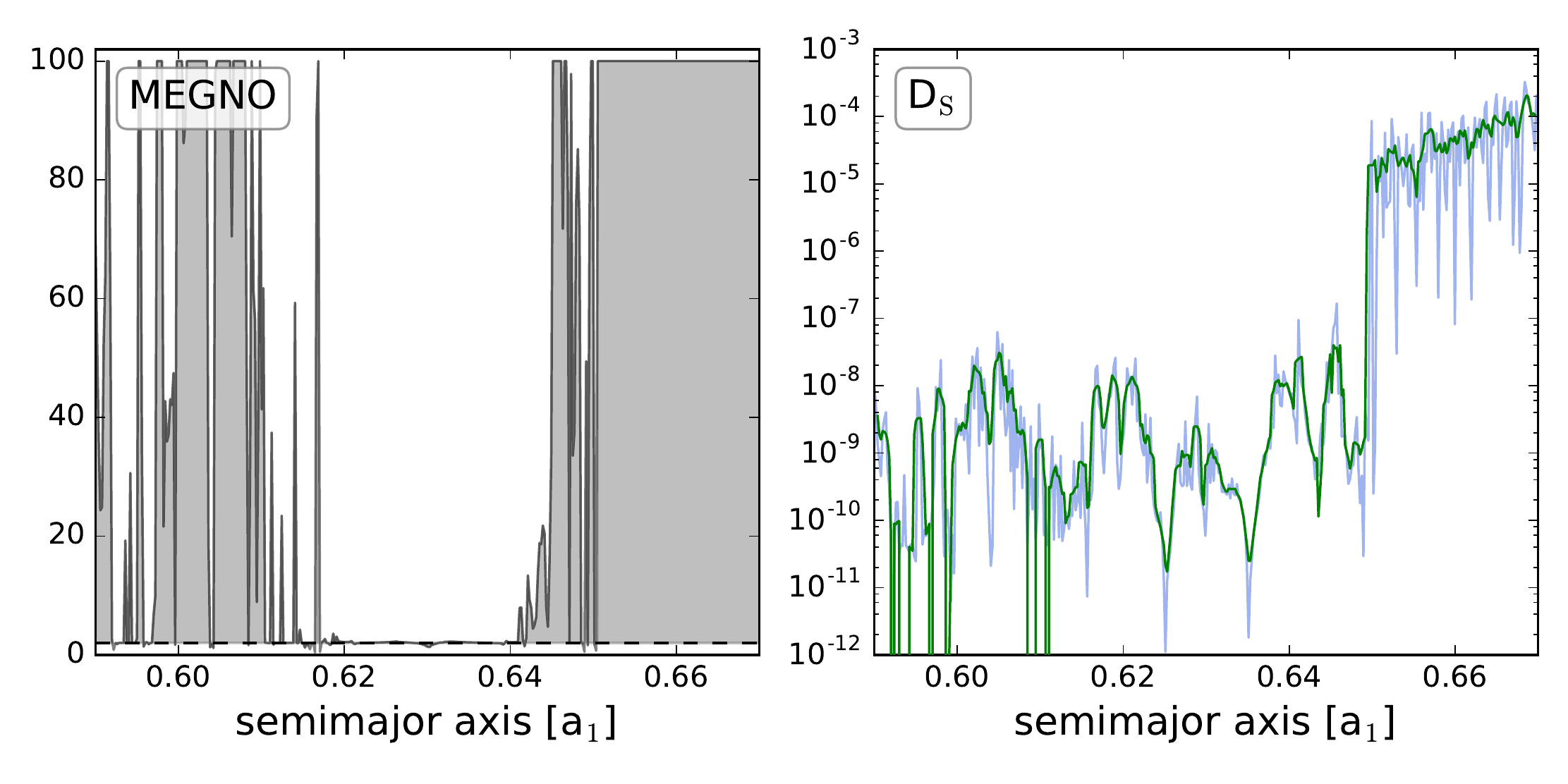}
\caption{{\bf Left:} MEGNO chaos indicator (cutoff value equal to 100) for a set of 400 
particles with initial conditions $a/a_1 \in [0.59,0.67]$ and $e=0.4$. Total integration time 
was $T=10^5$ years. {\bf Right}: Light blue lines show the diffusion coefficient $D_S$ estimated 
from Shannon entropy, while mean values (averaged over 5 neighboring points) are depicted in 
dark green.}
\label{fig8}
\end{figure}

The left-hand frame shows the value of the MEGNO chaos indicator at the end of the integration 
time. Values larger than $100$ were set at that limit. Recall that regular trajectories are 
associated with a MEGNO value about 2, indicated in the graph with a horizontal dashed line.
While initial conditions inside the 2/1 MMR libration domain, located between $a/a_1 \simeq 
0.62$ and $0.64$, appear regular, trajectories on either sides of the commensurability are 
dominated by chaotic motion. As was observed in Figure \ref{fig1}, the exterior region ($a/a_1 
> 0.64$) is characterized by the overlap of many high-order resonances. All initial conditions 
with $a/a_1 \gtrsim 0.65$ seem to form a single region of strong chaotic motion, with 
MEGNO reaching the maximum allowed value after only a few $10^2-10^3$ years. A different 
behavior is noted closer to the main separatrix, as well as for initial conditions with $a/a_1 
\lesssim 0.62$. This region does not appear to be dominated by a single connected chaotic sea, 
but by a series of smaller chaotic regions separated by more regular trajectories.

These differences in MEGNO among nearby trajectories define whether the chaos is local or 
global, and therefore affect the orbital instability. The right plot of Figure \ref{fig8} 
effectively shows that the region beyond $a/a_1 \gtrsim 0.65$ is associated to large and 
similar values of $D_S$, while much smaller values of the diffusion coefficient are obtained 
for the rest of the line segment. This indicates that $D_S$ does not necessarily correlate with 
the local value of the chaos indicator (e.g. MEGNO) even if the diffusion coefficient is 
calculated without the use of ensembles. It thus promises a more reliable indicator of global 
instability and not only of irregular motion.

\begin{figure}[t]
\includegraphics*[width=0.99\columnwidth]{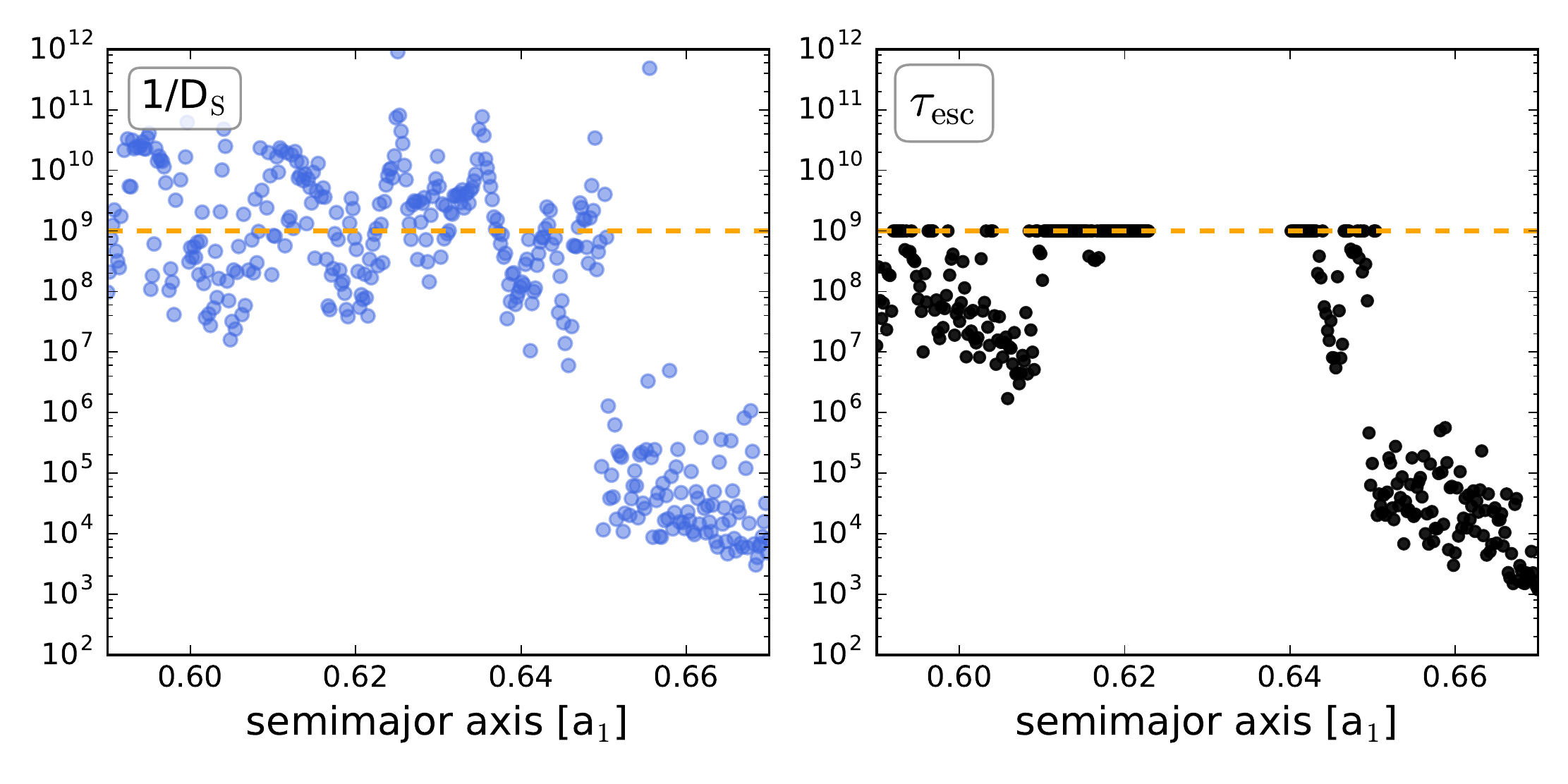}
\caption{{\bf Left:} Estimation of the escape time $\tau_{\rm esc} \simeq 1/D_S$ as deduced from 
the values of the unaveraged diffusion coefficients shown in the right-hand panel of Figure 
\ref{fig8}. {\bf Right}: Actual escape times, for the same initial conditions, obtained from an 
N-body simulation of the exact equations of motion. Total integration time was set at $T=10^9$ 
years (dashed horizontal orange line).}
\label{fig9}
\end{figure}

Figure \ref{fig9} now compares the predictions of the escapes times $\tau_{\rm esc} \simeq 
1/D_S$ (left), for the same 400 initial conditions with numerical integrations of the exact 
equations of motion (right) for a significant fraction of the initial conditions (those for 
which the black dots are depicted in the figure). In the N-body simulations the escape time was 
defined as the moment in which the trajectory satisfied any one of the following conditions: 
{\it (i)} eccentricity equal or larger than unity ($e \ge 1$), {\it (ii)} semimajor axis $a$ 
larger than twice that of the perturber, {\it (iii)} semimajor axis smaller than $0.1 a_1$, {\it 
(iv)} minimum approach to the planet closer than one-tenth of its Hill radius, or (v) a 
physical collision with either $m_0$ or $m_1$.

Total integration time for each initial condition was set to $T=10^9$ years (orbital periods of 
the primary). For comparison, if the system were to represent a perturber placed in the orbit 
of Jupiter, the time-span covered by the simulation would be larger than the age of the Solar 
system. Trajectories that remained bounded at the end of the integration are identified with 
black circles set at $t=10^9$ years. The orange horizontal dashed line in both plots aids helps
to visualize which initial conditions have instability times detectable with the simulations.

In so far integrations have proceeded, we find a very good overall agreement between $\tau_{\rm 
esc}$ and the N-body instability timescales, not only qualitatively but also quantitatively. 
The dispersion observed between close-by initial conditions is also very similar in both cases, 
lending credibility to our proposition that the differences in the values $D_S$ calculated with 
and without ensembles could be (at least partially) due to small-scale structures in the phase 
space.

In the outer circulation domain of the 2/1 MMR ($a \gtrsim 0.65$), the diffusion coefficients 
estimated with Shannon entropy yield instability timescales of the same order as those found 
from N-body simulations, even though these correspond to three orders of magnitudes the 
integrations times employed for the calculations of $D_S$ (i.e. $10^5$ years). Conversely, in 
the inner circulation domain ($a \lesssim 0.62$), the use of resurrected initial conditions has 
allowed us to correctly identify very short instability times, even extending the calculation 
of the entropy to the full $10^5$ years. Again, the N-body simulations shows a very good 
agreement with the model.

\section{Application to Other Configurations}

So far we have concentrated solely on the behavior of initial conditions in the vicinity of the 
2/1 mean-motion resonance, considering a perturber with mass similar to that of Saturn. In this 
section we will explore larger perturbers and other configurations.

\subsection{The 2/1 MMR with a Jupiter-mass Perturber ($m_1 = m_{\rm Jup}$)}

We begin increasing the perturbing body to a Jupiter mass ($m_1 = m_{\rm Jup}$) but preserving 
its orbit (i.e. $a=1 = 1$ AU and $e_1 = 0.05$). Since the extension of the 2/1 MMR is broader, 
we will consider initial conditions for the particles in the range $a \in [0.59,0.67]$ AU and 
eccentricities $e = 0.3$. The top left-hand frame of Figure \ref{fig10} shows, in broad red 
lines, the inner and outer separatrix of the 2/1 MMR. We will analyze the dynamical evolution 
of 400 particles with initial conditions $a \in [0.59,0.67]$ AU, as shown with the horizontal 
dashed black line. 

\begin{figure}[t]
\includegraphics*[width=0.99\columnwidth]{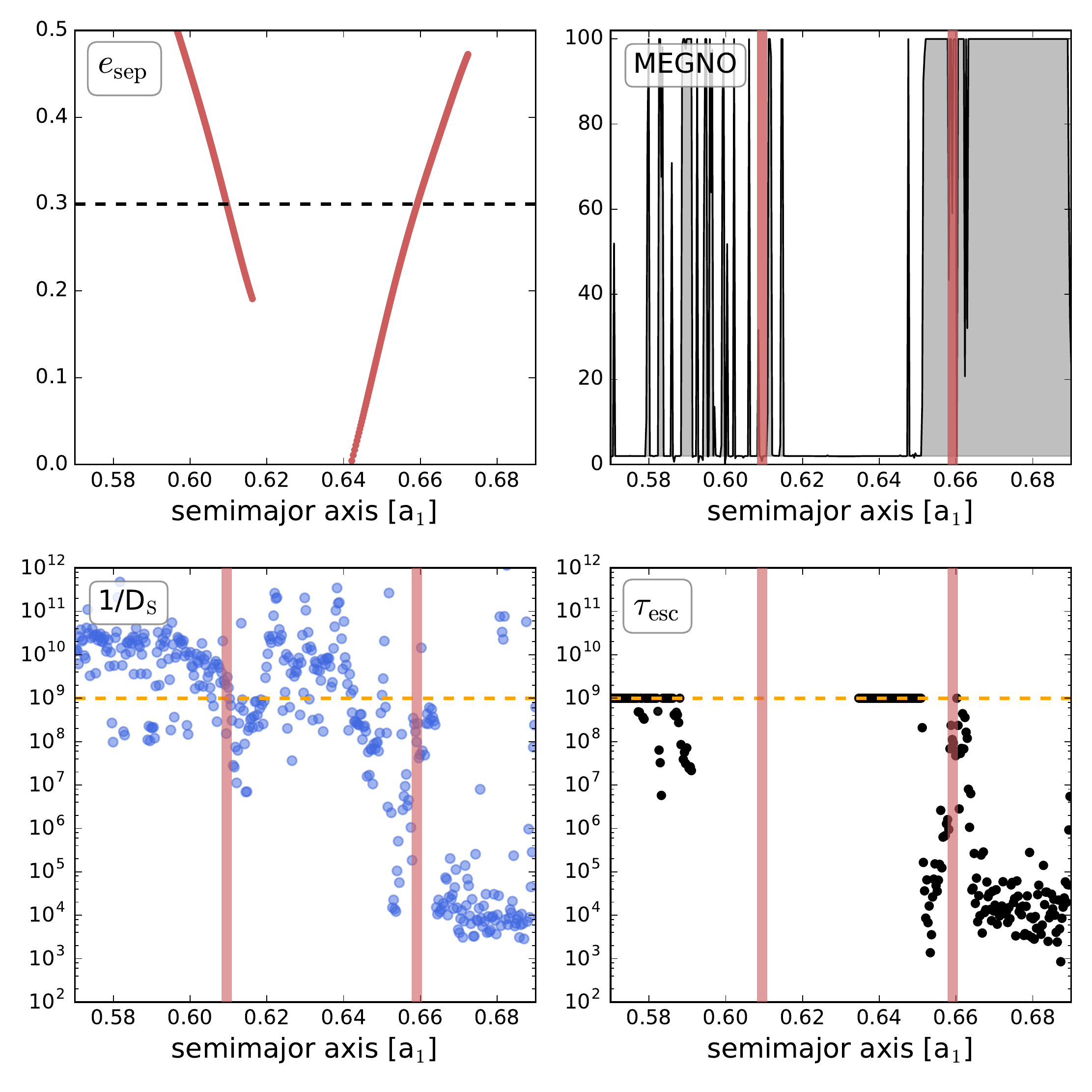}
\caption{{\bf Top Left:} Broad red curves show the outer and inner separatrix of the 2/1 MMR in 
the semimajor axis and eccentricity plane, considering all angles equal to zero. The horizontal 
dashed black line marks the location of the initial conditions analyzed in the other plots. {\bf 
Top Left:} MEGNO chaos indicator (cutoff value equal to 100) for a set of 400 particles with 
initial conditions with $e=0.3$. Total integration time was $T=10^5$ years. {\bf Bottom Left}: 
As with the left-hand plot in Figure \ref{fig9}, light blue circles show the estimated escape 
times $1/D_S$ calculated from the Shannon entropy. {\bf Bottom Right}: Actual instability times 
$\tau_{\rm esc}$ obtained from N-body simulations with integration time $T=10^9$ years.}
\label{fig10}
\end{figure}

The MEGNO chaos indicator over the total time-span is presented in the top right-hand graph. The 
broad light red vertical lines mark the location of the outer (left) and inner (right) branches 
of the separatrix. Comparing these results with those obtained for less massive perturbers (see 
Figure \ref{fig8}), we again obtain an extended chaotic sea beyond in the inner separatrix, but 
also a broader stochastic region inside branch of the separatrix itself. Conversely, the 
non-resonant domain inside of the outer separatrix now appears, in general, less chaotic and 
more interlaced with regular trajectories. It thus seems that, at least inward of the libration 
domain, the lower eccentricity of the initial conditions leads to more regular behavior, even 
with a larger perturbing mass.

The distribution of the inverse of the diffusion coefficients (i.e. $1/D_S$), shown in the 
lower left-hand frame, tell a similar story. Shannon entropy predicts very fast instability 
times, of the order of $\sim 10^3-10^4$ years, for initial conditions with $a \gtrsim 0.66$, 
and much longer ejection times on the other side of the main resonance. Inside the libration 
domain, most of the trajectories appear long-lived except for those in the vicinity of both 
branches of the separatrix.

The N-body simulations for some of the 400 initial conditions (bottom right-hand plot) appear 
very close to the predictions of our model. 

\begin{figure}[t]
\includegraphics*[width=0.99\columnwidth]{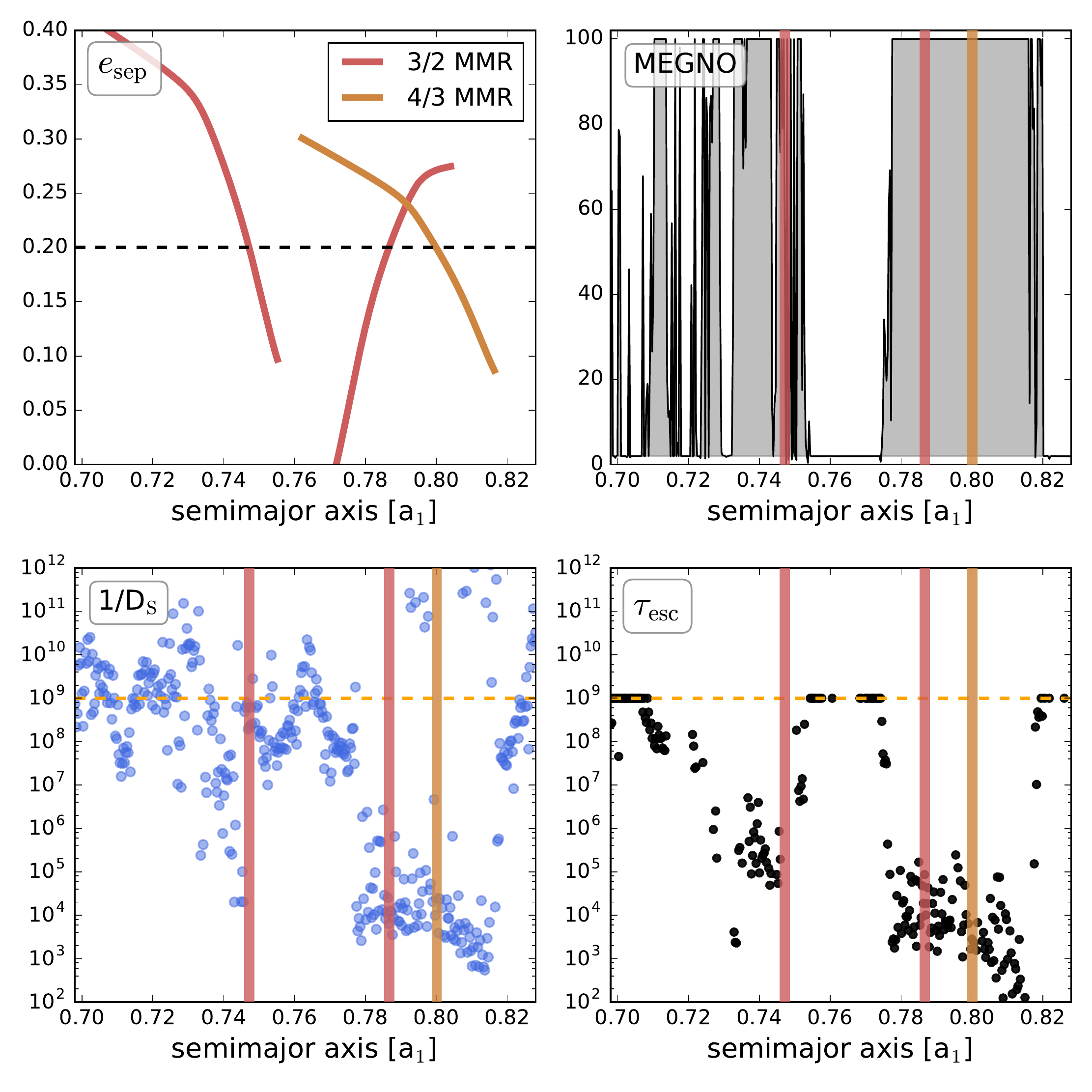}
\caption{Same as previous figures, but for initial conditions around the 3/2 MMR and adopting a 
perturber with mass $m_1 = 2.5 \times 10^{-4} \, m_0$.}
\label{fig11}
\end{figure}

\subsection{The 3/2 MMR with a Saturn-mass Perturber}

For the final application of Shannon entropy, we return to the smaller perturbing mass ($m_1 = 
2.5 \times 10^{-4} \, m_0$ but shift the set of particles closer to the perturber up to the 
vicinity of the 3/2 MMR. We thus again analyzed the dynamical evolution of 400 initial 
conditions, this time with eccentricities $e=0.2$ and semimajor axis in the interval $a \in 
[0.7,0.83]$. The top left-hand plot of Figure \ref{fig11} depicts this set as a black 
horizontal dashed line in the $(a,e)$ plane, while both branches of the separatrix of the 3/2 
MMR are shown in dark red. These curves were calculated using a semi-analytical model (e.g. 
Ramos et al. 2015) and results were truncated before reaching the collision curve. 

The outer edge of the interval of semimajor axis for the particles was chosen to reach the 4/3 
mean-motion resonance, whose separatrix is also included in the plot (light brown curve). It 
will prove interesting how this near-overlap will affect the instability of the trajectories in 
their neighborhood. The distribution of MEGNO, shown in the top right-hand frame, indicates that 
the this region is completely immersed in a chaotic sea, with the exception of initial condition 
beyond $a \gtrsim 0.82$ that seem located in a regular island deep within the 4/3 MMR. The same 
is observed inside the libration domain of the 3/2 resonance, while, again, below $a \lesssim 
0.73$ trajectories exhibit a complex alternation between chaotic and more regular motion.

The inverse of the diffusion coefficients, shown in the lower left-hand panel, translate 
chaoticity to estimations of the instability times. The region associated to the partial 
overlap between the 3/2 and 4/3 MMRs shows very fast escapes times, of the order of $10^4$ 
years, while much more stable trajectories are obtained inside both libration lobes. Finally, 
the non-resonant region inside of the 3/2 resonance shows intermediate values, with large 
instabilities close to the outer separatrix, while leading to longer escape times farther from 
the commensurability. 

The N-body simulations for some initial conditions are shown in the lower right-hand plot, 
and again the estimations of $\tau_{\rm esc}$ given by $1/D_S$ appear consistent 
with the predictions.

\section{Conclusions}

We have presented a series of applications of Shannon entropy as a numerical tool with which to 
estimate the diffusion coefficient and instability times of chaotic trajectories in the 
restricted three-body problem. We analyzed both resonant and non-resonant initial conditions in 
the vicinities of the 2/1 and 3/2 interior mean-motion resonances, and compared the results with 
direct N-body simulations.

In general, we have found very good agreements between our estimations and direct numerical 
results, even in cases where the actual instability times are several orders of magnitude longer 
than the integrations necessary for the entropy-based diffusion coefficients. 

More interesting is the fact that the computational effort to calculate the Shannon entropy is 
much smaller than the one required for estimating the diffusion rate by means of the variance of the action-like variables, since we have shown that ensembles of initial conditions are not 
necessary required. 

The results here obtained together with those presented in Giordano \& Cincotta (2018) suggest that the entropy would become an effective alternative to estimate a diffusion coefficient. In fact, a combination of different tools would provide a natural way to investigate the dynamics of 
dynamical systems. Indeed, the phase space structure could be displayed by some fast chaos 
indicator as the MEGNO or similar, that tells us about the location of resonances and their 
associated stable and unstable manifolds, chaotic regions and small stability domains embedded 
in the chaotic sea. But since chaos indicators are in principle unable to distinguish between 
stable or unstable chaos, the entropy could be computed in the chaotic domains in order to get 
an estimate of the time-scale of the unstable motion.

\begin{acknowledgements}
Most of the calculations necessary for this work were carried out with the computing facilities 
of IATE/UNC as well as in the High Performance Computing Center of the Universidad Nacional de 
C\'ordoba (CCAD-UNC). This research was funded by CONICET, Secyt/UNC and FONCYT. 
\end{acknowledgements}

\vspace*{0.5cm}
{\it The authors have no conflict of interest to declare.}

\end{document}